# The Haptic Paradigm in Education: Challenges and Case Studies

*Felix G. Hamza-Lup*
*Professor of Computer Science*
*Armstrong Atlantic State University*
*Felix.Hamza-Lup@armstrong.edu*

*Ioana A. Stanescu*
*Research & Development*
*Advanced Technology Systems, Romania*
*Ioana.Stanescu@ats.com.ro*

**Abstract**

*The process of learning involves interaction with the learning environment through our five senses (sight, hearing, touch, smell, and taste). Until recently, distance education focused only on the first two of those senses, visual and auditory. Internet-based learning environments are predominantly based on visual with auditory components. With the advent of haptic technology we can now simulate/generate forces and, as a result, the sense of touch. The gaming industry has pushed the "touch" on the "wire," allowing complex multimodal interactions online. In this article we provide a brief overview of the haptic technology evolution, potential for education, and existing challenges. We review recent data on the 21$^{st}$ century students' behaviors, and share our experiences in designing interactive haptic environments for education. From the "Community of Inquiry" framework perspective, we discuss the potential impact of haptic feedback on cognitive and social presence.*

**Introduction**

As defined in the Community of Inquiry (CoI) framework, social presence is the ability of learners to project their personal characteristics into the CoI, thereby presenting themselves as *real people*, and form meaningful connections with others to enhance collaborative learning experiences. CoI also explains cognitive presence as the extent to which learners are able to construct and confirm meaning through reflection and discourse in a four-stage process. This starts with a triggering event, then moves through exploration and integration, and culminates with the achievement of resolution (Garrison, Anderson & Archer, 2001).

From a distance learning environment perspective, haptic applications have the potential to significantly impact both types of presence discussed. There are two approaches for deployment of such applications:

- *Local deployment.* The required simulation components are downloaded and executed locally. Such an approach, while shielded from the network impairment, does not provide a direct interaction with the instructor.
- *Remote deployment.* The student and the instructor may *haptically* interact with each other. The sense of social presence is quantifiable in such cases, and is dependent on the network performance.

In what follows, we provide an overview of current technologies for creation of multimodal virtual environments, and a few related research efforts. We also discuss a case study that demonstrates the potential of haptic technology to affect the way students connect with each other and engage in the exploration of content.

## 1. Multimodal Virtual Environments as Learning Tools

A Multimodal Virtual Environment (MVE) provides multiple modalities to convey information. In the past the combination of audio and visual modalities made possible the production of illustrative explanations of various concepts, thus breaking the barriers of verbal communication.

Haptic, derived from the Greek *haphe*, means *pertaining to the sense of touch*. Another derivation is from the verb *haptesthai*, meaning *to touch*. Haptic technology is employed in the interface as input/output stimuli between the user and the computer. The communication is enabled via applied tactile and force feedback, or vibrations and motions in other words. The output stimuli can simulate the sensation of touching a virtual object. MVEs with visual, auditory, and, additionally, haptic stimuli can convey information more efficiently than with the use of a single sense, since the user manipulates and experiences the environment through multiple sensory channels. Each modality allows information exchange between the instructor and the student, establishing an experientially rich communication channel. Each communication channel, taken individually, has a specific communication bandwidth, dependent on the student capacity to use it. The bandwidth of the combination of these channels may be larger than their sum.

Kinesthetic learners, who make up about 15% of the population, struggle to fully grasp concepts by just reading or listening (Coffield, Moseley, Hall, Ecclestone, 2004). Without the haptic channel, the quality and amount of information conveyed through an interface is reduced, resulting in a narrower communication bandwidth and less efficiency during the learning process. This reduction negatively affects the cognitive results obtained through distance education and sets a barrier in adopting it on the same reliability basis as face-to-face education. As such, haptics may provide a medium to learn by doing, through first-person experience, enriching the environment presented through visual and auditory modalities.

## 2. Statistics on Students in the 21$^{st}$ Century

The current demographics of students, known as Millennials or Generation Y (Naomi Rockler-Gladen, 2006), have had an extensive exposure to vast arrays of technology, and are much better adapted to applying these technologies in learning environments than the previous generations. Bialeschki (2007) points to the rapid growth of technology as an aspect that must be considered in developing content for all learners, but particularly for the current generation, who have grown up with higher expectations of learning technologies. High multimedia expectations coupled with low attention spans increase the challenge of engaging a student in learning activities by conventional pedagogical methods. Cognitive studies have shown that students are more apt to learn if the method of exposure engages them. If students could apply their technological proficiency at a greater extent, to their learning objectives, they could more easily understand difficult and abstract concepts and better relate new information to what they already understand.

Data from 2009 indicates that the Internet usage has increased 362%, to almost 25% of the world population, in the last decade (Internet World Statistics, 2009). A recent study (Jones, 2009) shows that 30% of Internet users are Generation Y (born between 1977-1990). Moreover, according to Walsh and colleagues (2006), computer games are becoming an addiction for the new generation, as 45% of the most prolific gamers are 6 to 17 years old.

Surveys conducted in Europe show that 96% of all schools have Internet access today, and 67% of schools already have a broadband connection. Furthermore, 90% of classroom teachers use computers or Internet to prepare lessons, and 74% use computers as a teaching aid. Yet, the number vary widely, from the UK (96%) and Denmark (95%) to Greece (36%) and Latvia (35%) (New Europe, 2006).

One conclusion is that interactive technologies are relevant in drawing students' attention, and the incorporation of these technologies should be considered a key proposition in curriculum development. Specifically, the technologies that have the potential for influencing the largest number of learners should be of the highest priority.

## 3. Challenges: Haptic Data Distribution on the Internet

Today's computer networks have been designed to carry information that pertains primarily to two human senses: the auditory sense (e.g., sound and speech), and the visual sense (e.g., video, graphic, and text). Internet is now being reengineered to provide different levels of service for different types of traffic, such as supporting the transport of voice-over-IP protocol. Some current network architectures can support different Quality-of-Service (QoS) levels.

Allowing networks to carry information relating to other senses will open up an enormous potential for both new and dramatically improved applications. The ability to embed the sense of *touch* or *force feedback* into applications and then distribute them across the Internet may have a significant impact on distance learning and training. It is now a fact that the introduction of haptic components into interactive games has increased users' quality of experience, and, as a result, the market demand for these types of applications. As an example, Falcon Novint™ is an affordable haptic device for gaming that may reach millions of game players in a few years (Manhattan Scientifics, 2009).

Recent research (Yap, Marshall & Yu, 2007) has shown that each type of network impairment affects the force feedback characteristics in a specific way. The network delay can make the user feel a virtual object before it is visually in contact, or to penetrate solid objects that are assumed solid. Delay also desynchronizes the different remote copies of a virtual environment. Jitter gives the user the impression that the object's mass varies. Packet loss can reduce the amount of the force felt by the user. These impairments decrease the efficiency of multimodal distributed applications and may cause damage to the haptic device and to the end user.

To date, the network has not been seriously considered in the design of haptic compensation algorithms. However, the introduction of graded QoS architectures, such as Diffserv (Babiarz, Chan & Baker, 2006), into the next-generation Internet now offers the capability to limit the effects of packet delay, jitter, and loss.

### 4. Design of Efficient Haptic Simulations: Case Studies

Network impairments are an inherent negative characteristic of an *interactive* collaborative learning environment (Guttwin et. al, 2004). Due to such impairments, the integration of haptic interfaces with visual and auditory cues becomes even more challenging. Solutions were proposed to cope with delays for sensory modalities (visual and auditory), from a technical perspective. Additionally, from a perceptual perspective we must take into consideration the behavioral and neurological patterns specific to its human users. An efficient learning environment must provide an excellent *perceptual integration* (Stanney et. al, 2004), which is not only task-dependent, but possibly more difficult to attain than the technical integration. Discovering and defining simulators and training tools that would benefit form the haptic feedback is also difficult. One must identify the concepts that lend themselves best to such simulation, then design the learning experience and provide viable implementation alternatives. While the technical integration of the haptic sensation is important, so are the evaluation of technology acceptance, and the measurement of the technology's impact on learning.

### 4.1 HaptEK16 – Pascal's Principle and Hydraulics

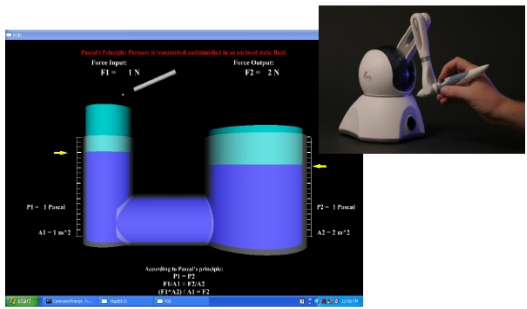

Figure 1: Student applies a force on one of the pistons to sense the pressure

Haptic Environments for K-16 (HaptEK16) is a novel visuo-haptic simulation for teaching physics concepts with emphasis on hydraulics (e.g., Pascal's principle). We have designed and implemented three activities to guide students from the simple concept of pressure (depending on two parameters: force and area) to the more complex system of a hydraulic car lift. The students can work through the activities at their own pace, interactively altering relevant parameters as they choose. Each activity is augmented in various ways by the haptic system, so that the students can apply and *feel* the forces in the experiments.

The force feedback is conveyed through a SensAble PHANTOM® Omni™ device. The system is developed using the Extensible 3D (X3D) modeling language (web3d.org) and the SenseGraphics™ H3D API. In the simulation, a student interacts with a 3D environment containing a set of interconnected pistons/pipes. A haptic stylus that is manipulated by the user to exert forces on the pistons (Figure 1) facilitates the interaction.

In the experiment the relationship between the force, surface, and pressure is explored. The student can change the surface and/or the amount of force applied and can *feel* as well as see the effects of these changes. The haptic component allows quantification of the forces and facilitates the student's understanding of the relationship among force, surface, and pressure while also demonstrating the incompressibility of liquids. The application had a strong impact on student engagement.

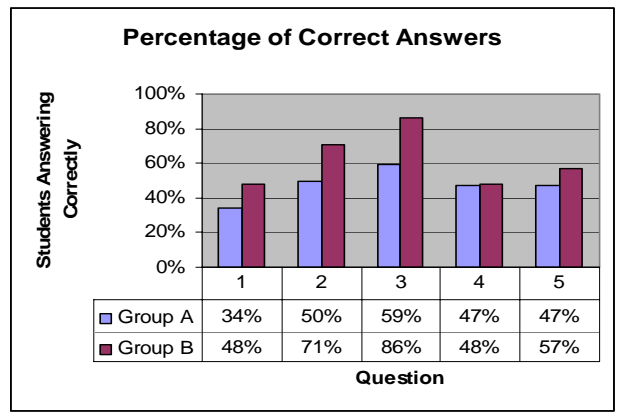

Figure 2: Test Results (Group A - no haptic; group B - with haptic) (Hamza-Lup, Adams 2008)

The assessment of the project was implemented in a classroom environment at a Richmond Hill high school in Southeast Georgia in 2007 and 2008. The results prove an increase in student's learning ability, as reflected in a set of pre- and post-tests (Hamza-Lup & Adams, 2008). Test scores (Figure 2) indicate that the students who received complementary instruction using the HaptEK16 simulator had better test scores and a higher level of engagement than the students with no haptic experience. A survey, administered at the same time, revealed that 94% of the students associated the simulator with a 3D game and expressed a strong interest to *play* with such simulations for other physics concepts. We believe that there is a strong relationship between playing and learning in this case, and cognitive learning can be significantly enhanced through play.

### 4.2 Haptics for Static and Dynamic Friction Simulation

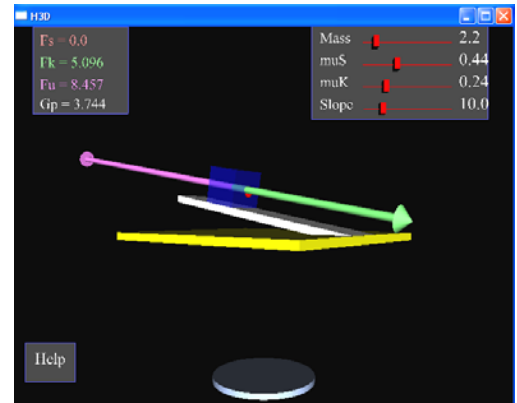

Figure 3: Users perceive the forces (direction and magnitude) while pushing the cube.

Fueled by the success of the HaptEK16 prototype, we decided to identify other concepts that would benefit from the introduction of the haptic paradigm.

Students learning about *friction* for the first time are confused both by its mathematical description as an inequality and by its nature as a force which varies based on the application of competing forces. Although a relatively simple experiment with an inclined wooden plane can demonstrate the principle of friction, we discovered several limitations to executing the experiment in a traditional approach:

- *Consistency of the experiment* is limited in the physical case. The inclined plane has a non-uniform surface, which makes the friction coefficient vary in different regions of the plane.
- *Customization of the experiment* is limited to the available materials in the lab; and extreme or interesting cases cannot always be realized. Students are required to fall back to the textbook and 2D illustrations.
- *User control over a continuous (large) range of physical parameters* is controllable by the user. Such a fine resolution cannot be achieved in a real experiment because of limited human mechanical ability.

Motivated by these limitations we designed and implemented an environment that simulates the force of friction and the associated paradigms. Students use the haptic device to manipulate a cube on an inclined plane and receive force feedback from the device (Figure 3). Students may apply varying amounts of force and directly receive varying resultant forces from the cube, they can also change the values that affect frictional force, such as the mass of the cube, the coefficients of static and kinetic friction and the slope of the plane along which the cube moves. The visuo-haptic simulation provides additional benefits, such as

- *Affordability*. Low-cost haptic devices that are connected to the existing computers in the school laboratories.
- *Portability*. The students can preview simulations online as part of a distance education tool, or in preparation for the lab.
- *Easy concept understanding*. Force vectors and their attributes can be visualized as 3D arrows. Such forces cannot be visualized in the traditional approach.

We are currently studying the efficiency of the simulator in a classroom setup. Preliminary results show a significant increase is students' engagement.

### 5. Conclusions

Practitioners and researchers have carried out studies to analyze the effect of haptic feedback on collaborative task performance, and to find out how haptic feedback can create a sense of social interaction within a collaborative virtual environment (CVE) (Kortum, 2008; Brewster & Murray-Smith, 2009; Hamza-Lup, Lambeth & LaPlant,

2009). The results suggest that basic haptic feedback increases the sense of social presence within the CVE (Hamza-Lup, Lambeth, & LaPlant, 2009; Laycock & Day, 2003). Utilizing haptics could improve the immersive experience of the user by adding the ability to not just perceive objects—visually and acoustically, as in current VE—but through tactile perception; this complement enables the users to virtually sense what they may be experiencing (Giannopoulos et. al, 2008). This ability should have a significant impact on the exploration dimension of the cognitive presence. Haptics can augment the *feeling of presence* in CVEs (Van Schaik, Turnbull, Van Wersch, & Drummond, 2004). Without remote deployment and assessment of haptic-based learning tools and environments, we can only assume the level of social presence, as defined by the CoI framework. Based on the results from CVEs, we believe that haptics will enhance the ability of learners to present themselves as *real people* in a distance learning environment, and to perform interactions with other students using a variety of sensory inputs, apart from text, video and audio. In addition, the ability to engage in richer tactile experiences may enhance the way in which students can explore content at a distance. As the cyberspace gains more and more features that improve the sense of reality, the quality of virtual exchanges of information increases to new levels and new demands, while students' comprehension expands to comprise fact-based learning.

It is important to remember that our goal is *not* the replacement of traditional learning tools that work well. We explore concepts and paradigms for which a visuo-haptic simulation will enable a better understanding among learners in a richer and more diverse environment. We envision such environments augmenting rather than replacing existing teaching methods. Finally, we are strongly convinced that there are a myriad of abstract paradigms that cannot be replicated via traditional experimentation, but would be better illustrated with the visuo-haptic approach.